\documentclass[pdflatex,sn-standardnature]{sn-jnl}

\jyear{2021}%

\theoremstyle{thmstyleone}%
\theoremstyle{thmstyletwo}%

\theoremstyle{thmstylethree}%

\begin{document}

\def\aj{AJ}               
\def\actaa{Acta Astron.}      
\def\araa{ARA\&A}             
\def\apj{ApJ}                 
\def\apjl{ApJL}                
\def\apjs{ApJS}               
\def\ao{Appl.~Opt.}           
\def\apss{Ap\&SS}             
\def\aap{A\&A}                
\def\aapr{A\&A~Rev.}          
\def\aaps{A\&AS}              
\def\mnras{MNRAS} 
\def\azh{AZh} 
\def\ssr{Space~Sci.~Rev.}
\def\nat{Nature}  
\def\pasj{PASJ}               
\def\pasp{PASP}
\def\na{New A}
\def\jcap{J. Cosmology Astropart. Phys.}
\def\baas{BAAS}

\title[Chemical and stellar properties of ETDs around the Milky Way]{Chemical and stellar properties of early-type dwarf galaxies around the Milky Way}


\author*[1,2]{\fnm{Vasily} \sur{Belokurov}}\email{vasily@ast.cam.ac.uk}
\author*[1]{\fnm{N. Wyn} \sur{Evans}}\email{nwe@ast.cam.ac.uk}
\affil[1]{\orgdiv{Institute of Astronomy}, \orgname{University of Cambridge}, \orgaddress{\street{Madingley Road}, \city{Cambridge}, \postcode{CB30HA}, \country{UK}}}

\affil[2]{\orgdiv{Center for Computational Astrophysics}, \orgname{Flatiron Institute}, \orgaddress{\street{162 5th Avenue}, \city{New York}, \postcode{10010}, \state{NY}, \country{USA}}}


\abstract{Early-type dwarfs (ETDs) are the end points of the evolution of low-mass galaxies whose gas supply has been extinguished. The cessation of star-formation lays bare the ancient stellar populations. A wealth of information is stored in the colours, magnitudes, metallicities and abundances of resolved stars of the dwarf spheroidal and ultra-faint galaxies around the Milky Way, allowing their chemistry and stellar populations to be studied in great detail. Here, we summarize our current understanding, which has advanced rapidly over the last decade thanks to the flourishing of large-scale astrometric, photometric and spectroscopic surveys. We emphasise that the primeval stellar populations in the ETDs provide a unique laboratory to study the physical conditions on small scales at epochs beyond $z=2$. We highlight the observed diversity of star-formation and chemical enrichment histories in nearby dwarfs. These data can not yet be fully deciphered to reveal the key processes in the dwarf evolution but the first successful attempts have been made to pin down the sites of heavy element production.}


\maketitle

\section{Introduction}\label{sec:intro}

Dwarf galaxies are by definition low in luminosity. The conventional threshold is fainter than $10^8 L_\odot$, corresponding roughly to the luminosity of the Small Magellanic Cloud \citep{Hodge71}. Their low luminosity is a sign of their struggle to make any appreciable amount of stars. Albeit scarce, these stars are nonetheless an invaluable source of information about physical conditions that may be exotic in the present-day Universe. The dwarfs' stellar populations reside in low-mass dark matter (DM) halos ($10^7 M_{\odot} \lesssim M_{\rm vir}\lesssim 10^{11}M_{\odot}$) \citep{Walker2009,Am11,Read2017,Jethwa18}, close to the threshold of star formation \citep[SF,][]{Shaye2004,Ryan-Weber2008}, often from nearly pristine gas \citep{Wise2012, Skillman2013}. 

In this review, we focus on studies of ETDs observed around the Milky Way. These comprise the dwarf spheroidals or dSphs (with stellar mass $M_\star \approx 10^{5-7} M_\odot$) and the ultrafaint dwarfs or UFDs ($M_\star \approx 10^{2-5} M_\odot$). Fig.~\ref{fig:sizeluminosity} shows the Milky Way satellites in the plane of size and luminosity. To give a feel for the bias which limits our view of the lowest luminosity ETDs to relatively small distances, objects within 50 kpc are coloured blue, beyond 50 kpc are coloured salmon. The ETDs occupy the domain fainter than $M_v=-15$, whereas the late-type satellites -- the Small and Large Magellanic Clouds -- are brighter. Ref.~\cite{Simon19} suggests the division between dSphs and UFDs is made in luminosity at $M_v = -7.7$. However, we argue that the boundary is more reasonably drawn in surface brightness. The dSphs have central surface brightnesses $\mu \lesssim$ 26 mag arcsec$^{-2}$, whereas the UFDs are all fainter, typically by 2-3 mag arcsec$^{-2}$. With this definition, the faintest known UFDs are the two feeble giants, Crater 2 and Antlia 2, which have surface brightnesses $\mu \approx 30$ mag arcsec$^{-2}$~\citep{Torrealba16,Torrealba19}. Note that Crater 2 and Antlia 2 would not be classified as UFDs following the definition based on luminosity. Minuscule surface brightness is truly a defining characteristic of UFDs as they routinely reach levels lower than those attributed to Low Surface Brightness galaxies (LSBs) and/or Ultra Diffuse Galaxies (UDGs) \citep[typically with $23<\mu<28$, see e.g.][]{Bothun1997,vanDokkum2015,Koda2015}. The dSphs have half-light radii between 170 pc (Leo II dSph) and 2660 pc (Sagittarius dSph), whereas the UFDs have half-light radii ranging from 24 pc (Segue 1) out to 2500 pc (Antlia 2). It is characteristic of ETDs that they are satellites of larger galaxies. About 50 ETDs are known around the Milky Way, with perhaps hundreds more within the virial radius, either unconfirmed or remaining to be discovered. 

The appeal of ETDs is two-fold. First, they represent the final stages of the complete evolution of a galaxy, thus offering a chance to benchmark our models of star formation, chemical evolution and feedback in relatively uncomplicated settings. Second, being the weakest amongst galaxies in the ability to retain gas, the ETDs host the least enriched stellar populations, opening the window into the earliest and most primitive stars. Beyond a megaparsec from the Sun, more numerous and representative samples of dwarfs are available, but none with such a detailed in-depth view of their denizen stars as provided by the resolved local ETDs. 

In the remainder of this section, we first sketch the life-story of the ETDs, emphasising their primitiveness and susceptibility to environmental effects. We compare their properties both to those recent interlopers, the Magellanic Clouds, as well as the destroyed dwarfs that built the stellar halo, such as the {\it Gaia} Sausage. Section 2 summarizes generic trends in the star formation histories and metallicities of ETDs, which bear the marks of their life-story. Finally Section 3 describes the evidence from abundances that pins down the early history of ETDs.

\subsection{Connection between early and late-type dwarfs}

ETDs are simple in appearance: they (usually) have spheroidal shapes and smooth stellar light distributions, lack large quantities of hydrogen and do not show any recent SF activity. They are antithetical to the late-type dwarfs (LTDs), that are gas rich and often have irregular shapes that bear signs of ongoing SF. The dichotomy of {\it late-type} and {\it early-type} dwarfs is thus, just like that of the more massive galaxies, to do with the distinction between {\it life}, i.e. active conversion of gas into stars and the associated chemical enrichment and feedback, and {\it death}, i.e. shutting down of SF and subsequent collisionless relaxation driven by gravity alone. The ETDs are simply the lifeless descendants of the LTDs \citep[][]{Kormendy85,Grebel03}. The intimate connection between the two types of dwarf galaxies is revealed by the continuity of their structural properties. This link is less perceptible around the Milky Way due to the dominance of quenched dwarfs, but is well established around more distant hosts and in galaxy agglomerations~\citep[][]{Boselli14,Venhola19,Carlsten21}.

The appearance of a galaxy that is star-forming (or alive) is determined, in the absence of strong environmental effects, by the state of its interstellar medium (ISM) and the mode of its SF~\citep[][]{Kauf07}. A bursty SF leads to irregularly shaped and clumpy galaxies, while steady SF rates are hosted by well-behaved discs \citep[][]{AK16,Dekel20}. Shallow potential wells of low-mass dwarf galaxies cannot easily contain highly energetic feedback processes like supernovae \citep[][]{MacLow99}, leading to erratic and bumpy SF histories \citep{Stinson07,Revaz09,Shen14,Muratov15}. At this mass scale, the entire matter distribution can be reshaped through non-adiabatic changes to the gravitational potential as a result of the bursty SF history \citep[][]{Read05,Mashchenko06,Pontzen12,Governato12,DiCintio14,Brooks14,Onorbe15}. With this in mind, most ETDs probably started off irregularly shaped, whilst the more massive ones were rotating discs~\citep{Klim09,Kaza11}. Their appearance today is the result of a morphological transformation unleashed by the environment they came to live in. Dwarfs are feeble and susceptible, they ultimately succumb to the effects of habitat.

Environmental effects eventually lead to the slowing down or switching off of the dwarf's SF. The earliest example of the interaction of a dwarf galaxy (or the proto-galaxy and its gas) with its surroundings happens during the epoch of cosmic reionization between redshifts of roughly 5 to 10 \citep{Rees86,Efsta92,Bullock00,Benson02,Ricotti05,Katz20}. Even if a dwarf survives reionization, its ability to form stars can be taken away easily. As benign an event as joining a group of other galaxies has far-reaching consequences. The dwarf is exposed to the group's hot corona \citep[][]{Gunn72} which, via ram pressure, removes the dwarf's gas supplies \citep[][]{Mayer06,Nichols11,Fill16,Simpson18}. After a while, in the absence of any fuel, the galaxy stops forming stars. Now a quenched satellite, the dwarf slowly relaxes and settles into a duller version of itself. 
However, the dwarf is now subject to the tidal forces of the group halo or its bigger neighbours. Tidal stripping, shocking and stirring lead to dramatic morphological transformations of dwarf galaxies~\citep[][]{Mayer01,Klim09,Kaza11}. Finally, mergers with other dwarfs provide a direct channel for scrambling the initial conditions~\citep[][]{Amorisco14,Rey19}, though merger rates are possibly too low to be important for the smallest objects \citep[][]{Fitts18}.

\subsection{Milky Way dwarfs were made early}

In view of this, it is not surprising that the majority of dwarf galaxies inside the Milky Way's virial radius ($\approx 200 $ kpc) have been transformed \citep[][]{Alan12}. In comparison with the two most massive satellites, the Large and Small Magellanic Clouds (LMC and SMC), the rest appear regular in shape with no prominent small-scale stellar substructure \citep{Martin08}, though the Sagittarius and Fornax dSphs do possess globular clusters. They have no detectable amount of hydrogen \citep[][]{Grcevich09,Spek14} and the last cinders of their SF have been extinguished \citep[][]{Weisz11,Weisz14,Weisz15}. The recent astrometric data from ESA's {\it Gaia} satellite delivered confirmation that these galaxies had {\it started their transformation early}. For a few satellites, some ambiguity in their orbits still exists due to our ignorance of the temporal evolution of the potential of the Milky Way at distances exceeding 100 kpc. However, most of the Galactic ETDs arrived to their current host many Gyrs ago \citep[][]{Helmi18orbits,Fritz18,Miyoshi20}. The Magellanic Clouds are a very recent accretion \citep[][]{Nitya06,Besla07,Piatek08,Nitya13}, which explains their irregular shapes, active SF and substantial gas reservoirs.

The accretion histories of the Milky Way satellites are intimately connected to their SFHs. Thanks to their proximity, their light distributions are resolved into individual stars. This permits construction of detailed colour-magnitude diagrams or CMDs with some moderate amount of mostly foreground contamination~\citep[early examples are presented in][]{Hodge71,Hoessel83,Freedman88,Tosi91}. Stellar distributions in the CMD can be modelled to infer the build-up of stellar mass as a function of age and metallicity \citep[][]{Tolstoy96,Dolphin97,Hernandez00,Gallart05}. This reveals that their stellar populations are dominated by ancient ($\gtrsim 10$ Gyr) and metal-deficient stars \citep[][]{Tolstoy09,Weisz11}. Some higher mass ETDs managed to hold on to some of their gas for a while and continued to form some stars at intermediate ages (4-8 Gyr), thus displaying complex SF histories \citep[][]{deBoerFornax}. Temporal resolution of the CMD-based SFH models deteriorates with look-back time as the spacing of the models in the CMD space starts to shrink. It is approximately $0.1-0.2$ Gyr at the age of 1 Gyr and dropping to $1-2$ Gyr for stars older than 8-10 Gyr \citep[][]{Hidalgo11}, though relative ages at the level of 0.5-1 Gyr are possible for ancient populations~\citep{Br12}. It is unfeasible to track short timescale variations in early SF activity predicted by the theoretical works discussed above. It is however possible to resolve later modulations in the SFH, typically associated with the in-fall and the ensuing interaction with the Milky Way \citep[][]{Miyoshi20,Rusakov21}.

The prevalence of ancient stellar populations in the ETDs presents a unique opportunity to study physical conditions that underpinned structure formation and chemical enrichment on small scales in the early Universe, i.e beyond $z=2$. Observationally, the combined effects of the DM halo growth regulated by the baryonic feedback are summarized by a relation between the galaxy's baryonic mass $M_*$ and the DM mass $M_{\rm vir}$ \citep[][]{Seljak00,Berlind02}. While the total stellar mass is available directly from observations, the total DM mass must be inferred via abundance matching or dynamical modelling \citep[][]{Vale04,Conroy09,Am11}. At the range of the halo masses occupied by ETGs, the relative strength of the baryonic regulation of galaxy formation reaches its maximum, leaving many of the low-mass DM halos completely devoid of stars \citep[][]{Bullock17,Wechsler18}. The marked decrease in the galaxy formation efficiency at low $M_{\rm vir}$ can be analysed to infer i) the steepness of the $M_* - M_{\rm vir}$ relation, ii) the amount of scatter in stellar mass at a given DM mass, and iii) the lowest DM halo mass to host a galaxy~\citep[][]{GK17,Jethwa18,Nadler19,Munshi21}. These phenomenological descriptors can then be linked to the underlying physics of the early Universe. For example, the change in the slope of $M_*-M_{\rm vir}$ curve can be connected to the impact of cosmic reionization. Much of the evolution of the $M_*-M_{\rm vir}$ relation at the low-mass end (e.g. the increased steepness and scatter at $M_{\rm vir}\lesssim 3 \times 10^9 M_{\odot}$) can be explained by the detrimental effects of the UV heating of the gaseous intergalactic medium during reionization~\citep[][]{Gnedin00,Okamoto08,GRUMPY}.

Reduced SF activity in the nearby ETDs can be exploited to understand the details of the synthesis and the delivery of heavy elements in supernova (SN) explosions and neutron star mergers (NSM). Compared to the field, it is easier to find a primitive star (i.e. one formed from the gas enriched by a small number of previous stellar generations) in a spectroscopic follow-up campaign \citep[][]{Starkenburg10,Frebel15}. More importantly, the chemical evolution of ETDs is straightforward to interpret. Small galaxies quickly finish accreting fresh gas \citep[][]{Fakhouri2010} and for the rest of their lives recycle the fuel acquired early. Thus, distributions of chemical abundances form tight and recognizable sequences that can be modelled. For instance, relative contributions of different kinds of polluters (including poorly constrained exotic types) can be inferred \citep[][]{Koba06,Koba20} and the yields of individual elements contributed by particular types of SNe pinned down \citep[][]{Kirby19,delosReyes20,Sanders21}. 

One example of such a pattern is the evolution of the ratio of the $\alpha$ element abundance (such as O, Mg, Si) to that of iron, Fe. As the ETD starts to form stars, [$\alpha$/Fe] values stay elevated (high plateau) but begin to decrease noticeably later in its evolution, as shown in the top panel of Fig.~\ref{fig:abundances}. This characteristic plateau+knee [$\alpha$/Fe] pattern is due to the change in relative contribution by core collapse supernovae (ccSN) and Type Ia SNe \citep[][]{Tinsley79,Gilmore91,McWilliam97,Venn04}. The two types of SNe show distinct [$\alpha$/Fe] behaviour due to i) differences in the mechanism of heavy element dispersal during the explosion (the heaviest elements end up locked in compact remnants after CCSNe, while much if not all~\citep{Kr13} the WD is destroyed during SN Ia) and ii) time delays between the SF and the SN explosion (massive stars, the progenitors of CCSNe collapse on the order of 10-20 Myr, while it takes on average $\sim$1 Gyr for a stellar binary to produce a WD that can explode as a SN Ia). 

For elements heavier than iron there exist two principal nucleosynthetic routes both requiring capture of neutrons onto heavy (e.g. Fe) seeds, but differing in neutron flux densities. The so-called {\it slow (s)}-process operates at low neutron densities while at high neutron densities, {\it rapid (r)}-process takes over \citep[][]{Burbidge1957}. The astrophysical site of the {\it s}-process has been identified with the AGB stars \citep[][]{Iben1983,  Gallino1998, Busso1999,Karakas2010}, while the exact origin of elements produced in the {\it r}-process is still debated~\citep[][]{Thielemann2020,Cowan2021}. It may take place in core-collapse supernovae~\citep[ccSNe][]{Hille1978,Takahashi1994,Wanajo2001,Farouqi2010,Curtis2019} including magneto-rotational explosions~\citep[][]{BK1970,Nishimura2017,Halevi2018}, or it may need
rarer but more efficient mergers of two neutron stars (NSs) (or a neutron star and a black hole)~\citep[][]{Lattimer1974,Eichler1989,Freib1999,Wanajo2014,Rosswog2014}.

\subsection{Ghosts of the dwarfs long gone}

The Milky Way's stellar halo is built from earlier accreted dwarf satellites. This provides a rich repository to investigate chemical and stellar population properties of small galactic systems long gone. With wide-area spectroscopic surveys and the astrometry from {\it Gaia}, it is now possible to unscramble the tidal wreckage and re-assemble the disrupted satellites.  Thanks to phase-mixing, the stellar debris can find its way close to the Earth. These bright stars are perfect targets for spectroscopic follow-up campaigns, offering an unprecedentedly high-resolution and high signal-to-noise view of their progenitor's chemical abundance patterns. The destroyed systems are not expected to be identical to the ETDs that survived to the present day. Nonetheless, the local ETDs are the obvious point of comparison because they too had their SF activity shut down a relatively long time ago. The destroyed dwarfs experienced SF cessation at different points further in the past, so on average their stellar populations should be more primitive. Spectroscopic studies of the stellar halo suggest it contains relatively more metal-poor stars than the massive, classical dSphs \citep[][]{Helmi06}, though this may in part be due to an incorrect metallicity analysis for stars below [Fe/H] $\sim -3$ \citep{Kirby2008}. The halo's metallicity distribution function is also more mixed \citep{Yo20} and it contains an unexpectedly large number of high [$\alpha$/Fe] abundance stars at high metallicity compared to the dSphs \citep{Shetrone98,Tolstoy03,Venn04}. Over the last 15 years, a vast population of low-luminosity, ultra-faint dwarfs (UFDs) was uncovered with photometric sky surveys~\citep[][]{Belokurov06,Belokurov07,Zucker06,Koposov15,DrlicaWagner15,Torrealba16,Torrealba19,Homma19}. Disrupted UFD systems may well be the principal contributors to the metal-poor end of the stellar halo \citep[see the discussion in ][]{Simon19}. At the metal-rich end, both the homogeneity and the higher SF efficiency (higher $\alpha$ abundances) of the halo stars are due to the dominance of the tidal debris of the ancient massive merger predicted by \cite{Deason13} and discovered using {\it Gaia} data \citep[the disrupted galaxy known as the {\it Gaia} Sausage (GS), or {\it Gaia} Enceladus,
  see][whilst \cite{Ev20} traces the history of the idea]{Belokurov18,Haywood18,Helmi18}.
  Stars in the GS are older than 10 Gyrs~\citep{Bo20}
and possess a chemical signature of low [Mg/Fe] and low [Ni/Fe]~\citep{Mont20}. The highest [$\alpha$/Fe] abundances in the local halo are contributed not by accreted satellites, but instead by the splashed prehistoric disc of the Milky Way itself \citep[][]{Nissen10,Gallart19,DiMatteo19,Belokurov20}.

\begin{figure}[h]
  \centering
  \includegraphics[width=0.9\textwidth]{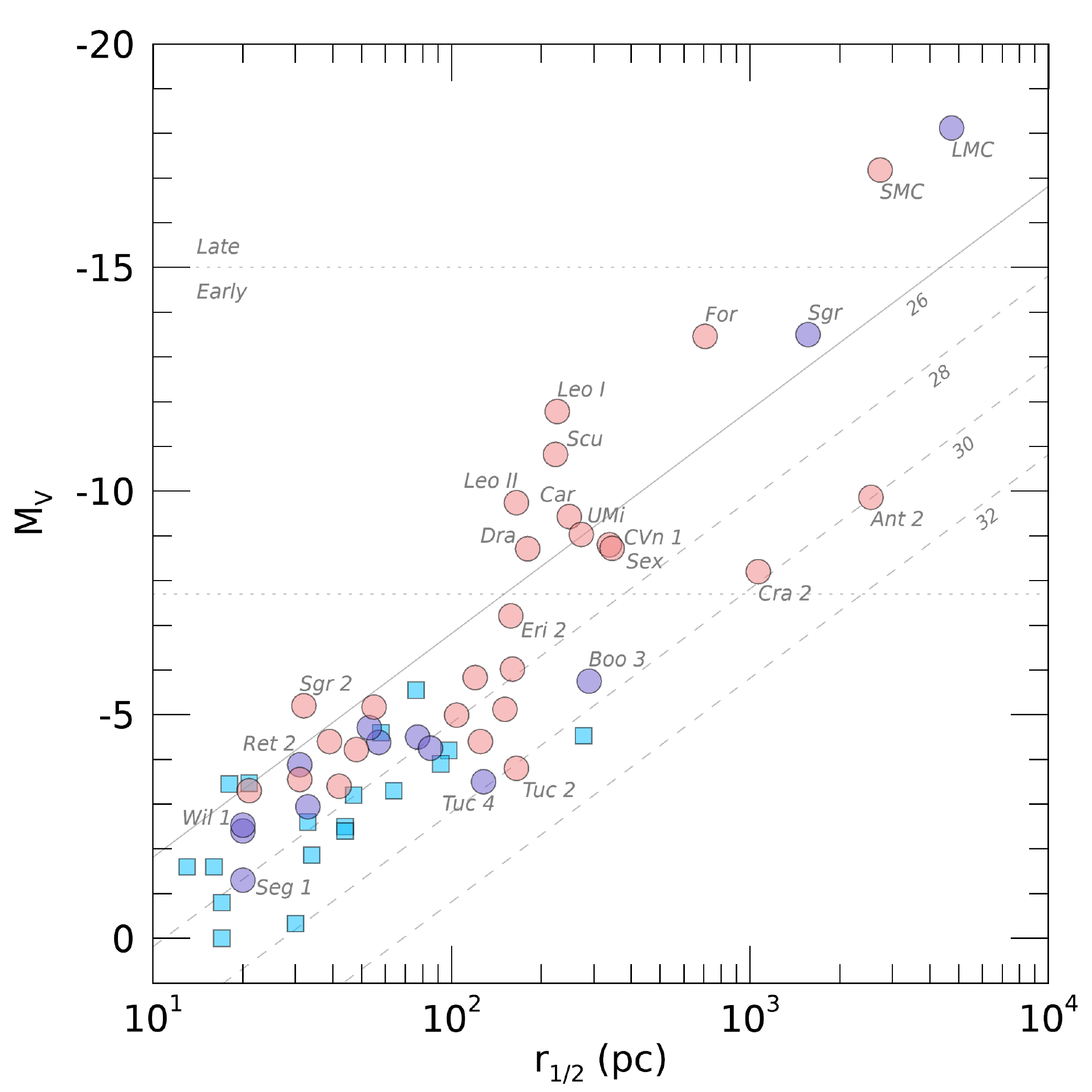}
  \caption[]{{\bf The size-luminosity relation for Milky Way satellites.} The half-light radius $r_{1/2}$ is shown against absolute magnitude $M_v$ using data from \cite{DW20} and \cite{Ji21}. The median relative size uncertainty is $\approx10\%$ while the median relative absolute magnitude uncertainty is $\approx5\%$ \citep[see the updated version of][]{Alan12}. Cyan squares show unconfirmed objects, while circles are confirmed and coloured blue if within 50 kpc, salmon if beyond. All of the Classical and some of the representative UF dwarfs are labeled. The diagonal lines mark 26, 28, 30 and 32 mag arcsec$^{-2}$. Ref.~\cite{Simon19} suggests that Classical dSphs lie above $M_v = -7.7$ (horizontal dotted line). We argue that Classical and UF dwarfs are better separated at constant surface brightness $\mu=26$ mag arcsec$^{-2}$ (solid diagonal line). Note that the classifications according to the two definitions above largely agree, excluding the unusually diffuse Crater 2 and Antlia 2. These two have luminosities similar to the faintest Classical dwarfs but an order of magnitude larger sizes.  The two late-types -- the Magellanics -- are comparable in size to Antlia 2, but $\approx 8$ magnitudes brighter. The {\it Legacy Survey of Space and Time} (LSST) will survey the uncharted domain of extreme surface brightness and large distances.}
   \label{fig:sizeluminosity}
\end{figure}

\section{Generic Trends in the Milky Way ETDs}

\subsection{Star Formation Histories}
\label{sec:SFH}

The star formation histories (SFHs) provide a chronology of how the Milky Way ETDs assembled their present-day stellar mass. The more massive dSphs usually possess multiple populations, differing in age, metallicity and kinematics. Their SFHs are bursty and episodic. For example, using CMDs built from deep {\it HST} photometry, Ref.~\cite{Rusakov21} showed that the Fornax dSph had a major SF episode at early times, a second strong burst about 5 Gyr ago and recent intermittent episodes from 2-0.2 Gyr ago. Mergers or interactions have been suggested as a probable cause of its spasmodic SFH~\citep{AmoriscoEvans12}. There is also evidence of tidally induced SF, as the intermediate-age and young SF events may correspond with pericentric passages of its orbit around the Milky Way~\citep{Pa22}. The Carina dSph provides another clean example of an episodic SFH. This is evident from its CMD, which shows at least three different main sequence turn-offs~\citep{Tolstoy09,Se11}. Carina had at least three major SF episodes, one at old times ($>8$ Gyr ago), a second at intermediate ages (4–6 Gyr ago), continuing into even more recent activity (2 Gyr ago). Using both wide-field photometry and spectroscopic data, \cite{deBoer2014} find that about 60 per cent of the stars in Carina formed in the episode at intermediate ages. The populations in Carina are substantially more mixed than Fornax -- as indicated by the small differences among their characteristic half-light radii or velocity dispersion~\citep{Ko16}. The Sculptor dSph presents a less extreme example of multiple populations~\citep{deBoerSculptor,Be19,Re22}. It experienced a single SF event limited to the first $\sim 1-2$ Gyrs after the Big Bang, producing about 70 per cent of its stars. However, it continued to form stars in its central parts after the epoch of reionization, leading to a centrally concentrated, metal-richer population. In the outer parts, the stars are overwhelmingly metal-poor and probably formed during the reionization epoch, or soon after its end. The kinematics of multiple populations in dSpsh helps constrain the structure of their DM haloes, specifically as to whether the central density is cored or cusped~\citep{Ev09,Wa11,Am12}.

The less massive dSphs do not possess multiple stellar populations. For example, Sextans presents an almost a flat [Fe/H] radial distribution from the spectroscopic measurements conducted by~\cite{Ki11}, indicating a SF burst shorter than $\sim 1$ Gyr. 
Using deep Suprime-Cam photometry, \cite{Be18} argued that the Sextans dSph stopped forming stars about 13 Gyr ago, close to the end of the reionization epoch. Likewise, the stellar population of Draco dSph is mainly old. Although some intermediate-age population is present in Draco, most of the SF (up to 90 percent) took place before $\sim 10$ Gyr ago, with no SF activity detectable in the last $\sim 2$ Gyr~\citep{Ap01}. In the Ursa Minor dSph, virtually all the stars having formed earlier than 10 Gyr ago and 90 percent having formed more than 13 Gyr ago ~\citep{Ca02}. Halting of SF in dSphs happens via three mechanisms, namely (i) cosmic re-ionisation, (ii) SNe feedback associated with the early SF epoch in the dwarf galaxy itself and (iii) ram-pressure stripping and tidal interactions with a nearby larger galaxy. The relative importance of each contribution is unclear.

The least massive of all the ETDs are the UFDs. They are almost wholly composed of very ancient stars.  Even Crater 2, one the most luminous UFDs, has relatively old populations on average, i.e. older than 10 Gyr, while showing signs of somewhat extended SF  \citep[][]{Walker2019}. Typically, the UFDs have SFHs where $\gtrsim 75$ percent of the stars formed by $z \sim 10$ (or 13.3 Gyr ago) and 100 per cent by $z \sim 3$~\citep{Br14}. This is consistent with the picture that SF in these smallest dark-matter sub-halos was suppressed by the reionization of the universe. Some differences do appear to be present amongst the UFD population. For example, UFDs associated with the Magellanic Clouds show quenching times about 600 Myr more recent than those of other Milky Way UFDs, though the differences are probably within the errors~\citep{Sa21}. Eridanus II is a UFD located close to the Milky Way's virial radius. In the discovery paper~\citep{Koposov15}, there were what turned out to be misleading hints of young or intermediate age populations. With deeper ACS/HST CMDs reaching the oldest main sequence turn-off (MSTO), we now know that Eri II formed the bulk of its stars in a very early and extremely short ($<500$ Myr) burst~\citep{Gal_Eridanus}. By estimating the number of SNe events and the corresponding energy injected into the ISM, Eri II could have been quenched by SNe feedback alone.

\subsection{Metallicities}
\label{sec:metallicity}

Chemical properties of ETDs go hand in hand with their SF activity: the larger accumulated stellar masses (more extended SF histories) correspond to higher average metallicities, (computed using a representative sample of stars in the galaxy). 

\begin{figure}
  \centering
  \includegraphics[width=0.9\textwidth]{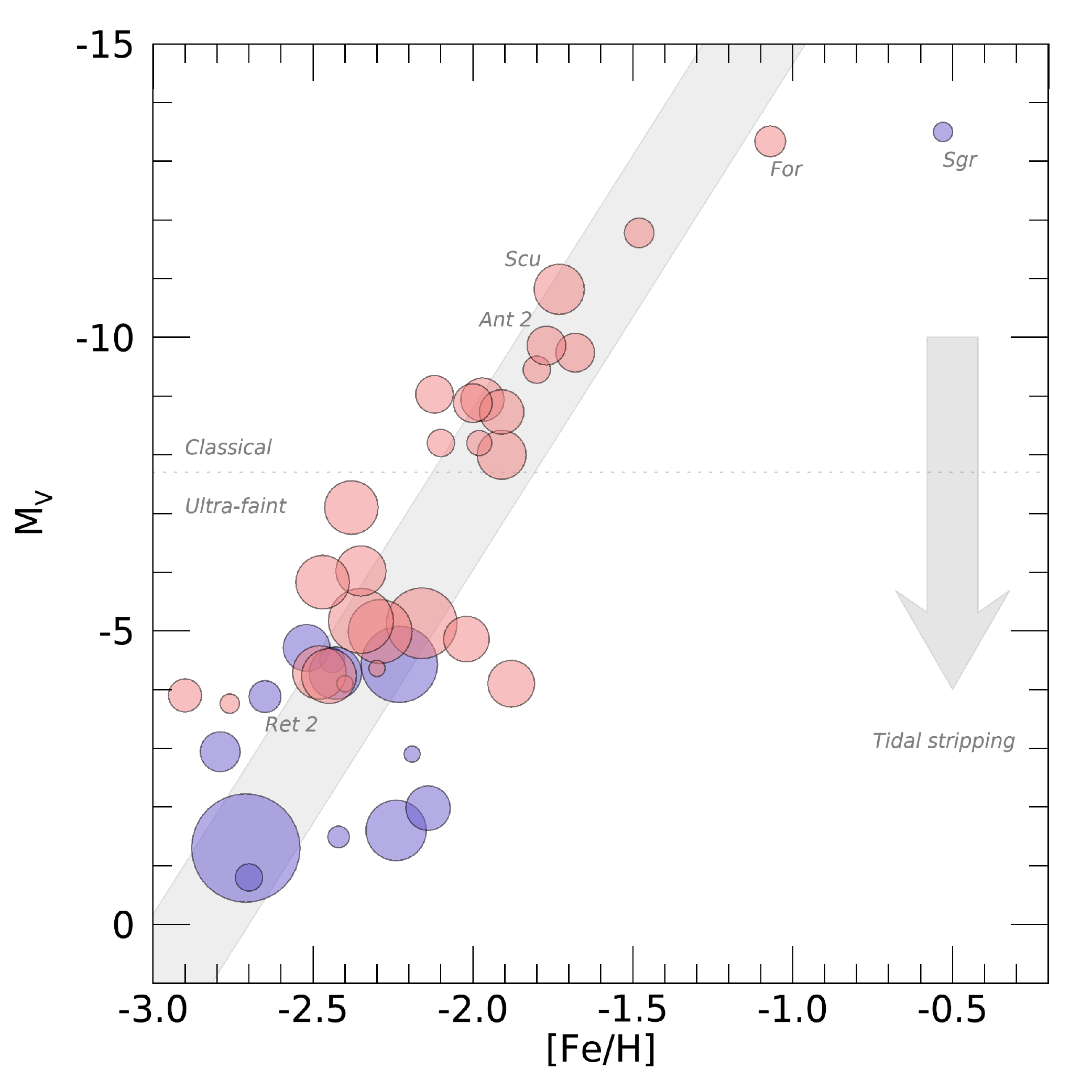}
  \caption[]{{\bf The (stellar) mass-metallicity relation for Milky Way satellites}, with the same colour-coding as in Fig~\ref{fig:sizeluminosity} using data from \cite{Simon19}. The size of the circle is proportional to the width of the metallicity distribution. Tidally stripped satellites evolve downwards as shown by the arrow. The grey band shows the fit to all Local Group dwarfs proposed by \cite{Kirby2013}, which holds over a range of $\gtrsim 10$ magnitudes.
  }
   \label{fig:massmetallicity}
\end{figure}

A correlation exists between the stellar mass or luminosity of a dwarf galaxy and its mean metallicity. For the Milky Way satellites, the data are shown in Fig~\ref{fig:massmetallicity}, together with the empirical fit of \cite{Kirby2013}. In fact, the mass-metallicity relation is well established for both gas-phase \citep[][]{Tremonti2004,Kewley2008,Mannucci2010} and stellar \citep[][]{Gallazzi2005,Zahid2017} abundance estimates across a wide range of galaxy masses. The shape of the correlation is close to linear for galaxies with masses lower than $10^{10.5}$, but starts to flatten beyond that. The evolution of the shape as a function of galaxy mass indicates that there are multiple competing physical processes at play (such as gas and metal flows in and out of the galaxy). 

Both ETDs and LTDs around the Milky Way lie on the same linear mass-metallicity relation \citep[][]{Kirby2013}. Over recent years, the range of masses $10^3<M/M_{\odot}<10^9$ has been enlarged by the UFDs, broadening the metallicity spread at fixed mass and extending the range to even fainter systems with $\sim10^2 M_{\odot}$. There is evidence of a floor in mean stellar metallicity in the low mass regime~\citep{Simon19}. While the increased metallicity spread can be accounted for by the differences in the amount of tidal stripping the galaxy can incur at fixed observed stellar mass, the flattening of the mass-metallicity relation at low masses is not fully reproduced in current numerical simulations \citep[][]{Jeon2017,Revaz2018,Wheeler2019,Applebaum2021,Grand2021}.

Characterizing an entire galaxy with an average metallicity value is an over-simplification. Even though ETDs and LTDs occupy the same continuous mass-metallicity relation, in detail, their individual [Fe/H] distributions are distinct~\citep{Kirby2013}. Metallicity distribution functions (MDFs) of LTDs are consistent with ongoing SF activity, while the MDFs of the brightest ETDs show a sharp cut-off at high metallicity, indicative of abrupt removal of their gas and rapid shut-down of their star formation. These MDF differences may pose a problem to the current theory in which dSphs are simply LTDs transformed by their interaction with the Milky Way. To understand the roles feedback and ram-pressure play, it is crucial to reconstruct and compare their SF and orbital histories \citep[][]{Miyoshi20}. Even after the complete shut-down of SF, the dwarf's past SF rate can be gleaned from the metallicity of the $\alpha$-knee (see Figure~\ref{fig:abundances}), which indicates how much self-enrichment has happened on the timescale corresponding to the onset of SNe Ia. Amongst the classical dSphs, the satellite's luminosity (and therefore total stellar mass) correlates reasonably well with the metallicity of the $\alpha$-knee \citep[e.g.][]{deboer_sgr2014,Kirby19, Rei20}, but this pattern is broken for the two largest satellites, the LMC and the SMC \citep{Nidever2020}. In these massive dwarfs, the contribution of SN Ia is shown to dominate Fe production at metallicites much lower than expected.  Using the most recent APOGEE data \cite{Hasselquiest2021} not only confirm the results of \cite{Nidever2020} but also reveal that the LMC and SMC both experienced a late uptick in $\alpha$-abundances symptomatic of a recent burst of star formation. The early SF inefficiency and the subsequent SF activity may be linked with the particular orbital evolution of the Clouds, i.e. residing in a distant low-density environment before falling into the Milky Way recently. Another massive satellite galaxy currently in a tidal interaction with the Milky Way, the Sgr dwarf, shows a more metal-rich $\alpha$-knee and is consequently inferred to have had a stronger SF than either of the Clouds \citep[see][]{deboer_sgr2014,deboer_sgr2015,Hasselquiest2021}.

\section{Nucleosynthetic Trends}
\label{sec:chemistry}

The mutation of LTDs into ETDs is accompanied by gas loss and cessation of star formation. Proximity to a large galaxy around the time of formation leads to ram pressure stripping of the dwarf's gas, truncating its SF activity. Ref.~\cite{Gallart15} used SF histories of local dwarfs to propose two types of dSphs: slow and fast. The fast ones are born in high density environments and form their stars early, while the slow ones stay away from big galaxy concentrations and thus extend their SF histories. Classification into fast and slow dwarfs is, however, rather ambiguous. Both Fornax and Sagittarius started with a SF burst, but continued to form stars later on, despite residing not too far from the MW \citep[see][]{Hasselquiest2021}. Note that the ram pressure is not the only means of unbinding the gas from the dwarf, it can also be done with feedback from SN explosions \citep[see][]{Revaz09,Sawala2010}.

In the last decade, progress in assembling large homogeneous spectroscopic samples \citep[e.g.][]{Kirby2011_alpha} has been matched by efforts to develop flexible chemical evolution  models to interpret the data \citep[][]{Andrews2017,Weinberg17,Spitoni2017}. According to the models, the gas outflow rate is discernible from the evolution of the $\alpha$-abundances with metallicity: stronger gas ejection makes the knee steeper \citep[see Fig 3 of][]{Andrews2017}. 
 
However, observational evidence for knees in $\alpha$-element trends with metallicity is mixed. For instance, analysing eight of the Milky Way dSphs, Ref.~\cite{Kirby2011_alpha} find no strong evidence for a knee in a sample limited to [Fe/H]${>}-2.5$. All eight galaxies exhibit similar mean tracks in all four $\alpha$-elements considered. No clear plateaus are visible, though most tracks do show change in steepness at intermediate [Fe/H]. In a follow-up study, \cite{Kirby19} detect hints of $\alpha$-knees in several dwarfs, all at low metallicity, [Fe/H]$<-2$. Unambiguously, dwarfs fail to hold on to the metals they make even if the physics of gas ejection are poorly understood \citep[see][]{Kirby2011_metals}.

To date, it has been securely established that the dwarf's chemical evolution largely depends on its stellar mass \citep[see e.g.][]{Kirby2011_alpha,Escala2018, Rei20}, but the details of gas inflow and outflow remain muddled by the uncertainties in the theoretical yields and SN rates, as well as the physics of the gas ejection.  This inspires much recent efforts exploiting medium and high resolution spectroscopy to constrain the properties of the stellar explosions that govern chemical evolution of galaxies. Motivated by this, we focus on chemical results emerging from fresh high-resolution studies aimed at pinning down the very sources of the heavy element production.

\subsection{Extremely metal-poor stars}

The lowest metallicity stars are unique tracers of the first bouts of chemical enrichment in the young Universe \citep[][]{Umeda2003,Heger2010} and as such have been vigorously hunted down both in the Milky Way and its satellites~\citep[][]{Ryan1996,Cayrel2004,Venn04,Beers2005,Frebel15}. Using the demarcation suggested by \cite{Beers2005}, extremely metal poor (EMP) stars are those with the iron abundances below $1/1000$ of the Solar, or [Fe/H]$<$-3, i.e. similar to the metallicities of metal-poor Damped Lyman-$\alpha$ systems \citep[][]{Pettini2008,Cooke2011}.  Early studies of EMPs revealed that at low metallicities, the fraction of the carbon-enhanced metal-poor (CEMP) stars dramatically increased~\citep[][]{Beers2005,Aoki2007,Yong2013,Sa15,Yoon2016}. CEMP stars come in a number of flavours. CEMP-s (or CEMP-r/s) stars exhibit over-abundance of s process (or r and s process) elements, respectively. They make up around 80 \% of all CEMP stars~\citep{Aoki2007}. The favored production channel of CEMP-s stars is mass transfer of carbon and s-process material from the envelope of an asymptotic giant-branch star to its (presently observed) binary companion~\citep{Sn08}. CEMP-r/s stars originate in their natal gas clouds enriched by early SNe~\citep{Ha11}. CEMP-no stars exhibit no strong neutron-capture-element enhancements and are preferentially found at the lowest metallicities~\citep{Aoki2007}. Possible progenitors for CEMP-no stars include massive  stars with [Fe/H] $< -6$, which models suggest have greatly enhanced abundances of C, N and O~\citep{Meynet2010,Norris2013}. 

The EMP/CEMP stars are linked to the bigger question of the properties of Population III stars, i.e. those first, presumably metal-free, objects that seeded structure formation and chemical evolution in the Universe \citep[][]{Abel2002,Bromm2011,Stacy2014}. While the CEMP stars observed today are not representatives of this primordial population, some of them may be Population III's direct descendants.

A number of EMP stars have been identified in the classical dSphs, such as Sculptor, Fornax and Sagittarius~\citep[][]{Tafelmeyer2010,Fr10,Chiti2018,Hansen2018,Chiti2020,Yoon2020}. Yet, the low-[Fe/H] wing of the distribution in ETDs is still weaker than that of the Galactic halo. Two recent developments have helped to reconcile the MDFs of the field and the satellite stars. First, the UFDs exhibit significant metallicity spreads and their mean iron abundances are an order of magnitude lower than the classical dSphs. As is evident from their SFHs, the UFDs contain a larger proportion of metal-poor and EMP stars~\citep[][]{Frebel15,Yoon2019}.  While carbon production channels in the early Universe operate in all galactic environments (stars with high C abundance are found in the Milky Way, dSphs and UFDs), the fraction of CEMP stars in UFDs is nearly 10 times higher compared to the dSphs \citep[][]{Norris2010Segue,Norris2013,Yoon2019}. Secondly, using {\it Gaia} DR2 proper motions, \cite{Sestito2019,Sestito2020} demonstrated that the orbits of over a quarter of all field stars with [Fe/H]$<$-4 are prograde and concentrated within 3 kpc of the Galactic plane. One explanations is that these ultra metal-poor stars were formed in the early disc and were kicked up by subsequent mergers~\citep[][]{Villalobos2008,Haywood18,Belokurov20}. Alternatively, according to \cite{BK2022}, these EMP stars represent a kinematically biased sample of stars born in the ancient hot, messy, pre-disc Milky Way characterised by a highly stochastic star formation and chemical enrichment.

Not all of the most metal-poor stars in the Milky Way are carbon-rich. A handful of metal-poor, but carbon-normal, stars have been discovered in the halo~\citep[][]{Caffau2011,Norris2013,Star2018} probing formation channels in the early Universe distinct from their C-rich counterparts. The lowest carbon abundance metal-poor star was discovered by \cite{Skul2021} in the Sculptor dSph. This star is surprisingly different from the bulk of the C-normal stars in the halo: at [Fe/H]$\approx$-4, it shows a pronouncedly non-uniform $\alpha$-element abundance composition. For example, at low [Mg/Fe], high [Ca/Fe] and [Ti/Fe] are observed. \cite{Skul2021} compare their measured abundances of C, Na, Mg, Al, Si, Ca, Sc, Ti, Cr, Mn, Co and Ni to theoretical models and determine that the best-fit is produced by an enrichment pattern seeded in a hypernova explosion of a $20 M_{\odot}$ zero-metallicity star. This is one of the best candidates for a descendant of a massive, metal-free Population III object.

\subsection{Progenitors of SN Ia explosions}

Type Ia SNe have been used to demonstrate that the expansion of the Universe is accelerating~\citep[][]{Riess98,Perlmutter99}. Notwithstanding this, the physics behind their use as standard candles remains obscure, as does the exact make-up of the progenitors. Both leading scenarios of SN Ia rely on formation of a tight binary where at least one of the companions is a stripped nucleus of an evolved star, i.e. a white dwarf \citep[][]{Whelan73,Iben84,Webbink84}. It is the WD that eventually explodes after either accreting enough material from its neighbour, or merging. The principal difference in the explosion theories then lies with the second object in the binary. If the donor of the fuel for the explosion is a normal non-degenerate star, it is postulated to fill its Roche lobe and transfer its (usually) H-rich gas to the companion until the accreting WD reaches the Chandrasekhar limit of $\sim1.4M_{\odot}$ and explodes \citep[][]{Whelan73} (though non-degenerate stars might be able to donate He to trigger a Type Ia SNe, see~\cite{Ib87}). On the other hand, the donor object could be a WD itself. Then, relatively small amounts of accreted He are required to detonate the explosion in the WD core, thus resulting in the total progenitor mass ranging from sub-Solar to approximately Chandrasekhar \citep[see e.g.][]{Woosley94,Bildsten07,Pakmor13,Shen14SN,Shen18}. Differences in the progenitor central densities lead to noticeable variation in the rate of production of heavy elements. In particular, Mn and Ni (but also Cr and C) are sensitive tracers of the SN Ia progenitor mass~\citep[][]{Seit13}.

In local ETDs, it is possible to unpick the contributions of Type II and Type Ia SNe to the chemical composition of the ISM in its early state. The comparison of the abundances of individual {\it light} elements against those in the {\it iron-peak} group can provide tight constraints on the yields of SN Ia and thus shed light on the nature of its progenitor. In some extraordinary cases, the analysis can be carried out for a single
star. \cite{McWilliam18} demonstrate that a relatively metal-rich star in the Ursa Major dSph exhibits low [$\alpha$/Fe] ratio as probed by Mg, Si, Ca, Ti and indicative of a prominent Type Ia contribution, while simultaneously possessing only traces of Sc, Mn, Cu, V, Ni and Zn. Comparing their measurements with theoretical yield predictions, \cite{McWilliam18} argue that such a drastic under-production of the iron-peak group elements can only be explained if this star was born from the gas enriched by a sub-Chandrasekhar supernova. Phenomenological descriptions of chemical evolution have been developed to explain the behaviour of several elements with increasing metallicity for a number of dSphs~\citep{Kirby19,delosReyes20}. While sub-Chandrasekhar explosions must have dominated the enrichment of Sculptor, the more metal-rich systems, such as the Fornax and Leo I dSphs, have SN progenitors with near-Chandrasekhar masses.

This is in agreement with the conclusions of \cite{Sanders21} who employ multi-source models similar to those described in \cite{Weinberg17} and \cite{Spitoni2017} to study chemical enrichment histories of several massive dwarfs including the LMC, the SMC, Sculptor, Sgr, the GS progenitor, as well as the MW Bulge. Comparing to the most recent models of double-degenerate SN explosions \citep[such as those by e.g.][]{Gronow21,Gronow21met}, \cite{Sanders21}
conclude that sub-Chandrasekhar supernovae have contributed
significantly in all systems studied. Depending on the model
considered, at low metallicities, sub-Chandrasekhar SNe are
likely responsible for between 60\% and 100\% of all Type Ia
explosions, at the intermediate metallicities the variations in the [Mn/Fe] yield could be a sign of changing progenitor mass, while at higher [Fe/H] the Chandrasekhar channel may become increasingly more important.

\subsection{The nature of the {\it r}-process sites}

\begin{figure*}
  \centering
  \includegraphics[width=0.99\textwidth]{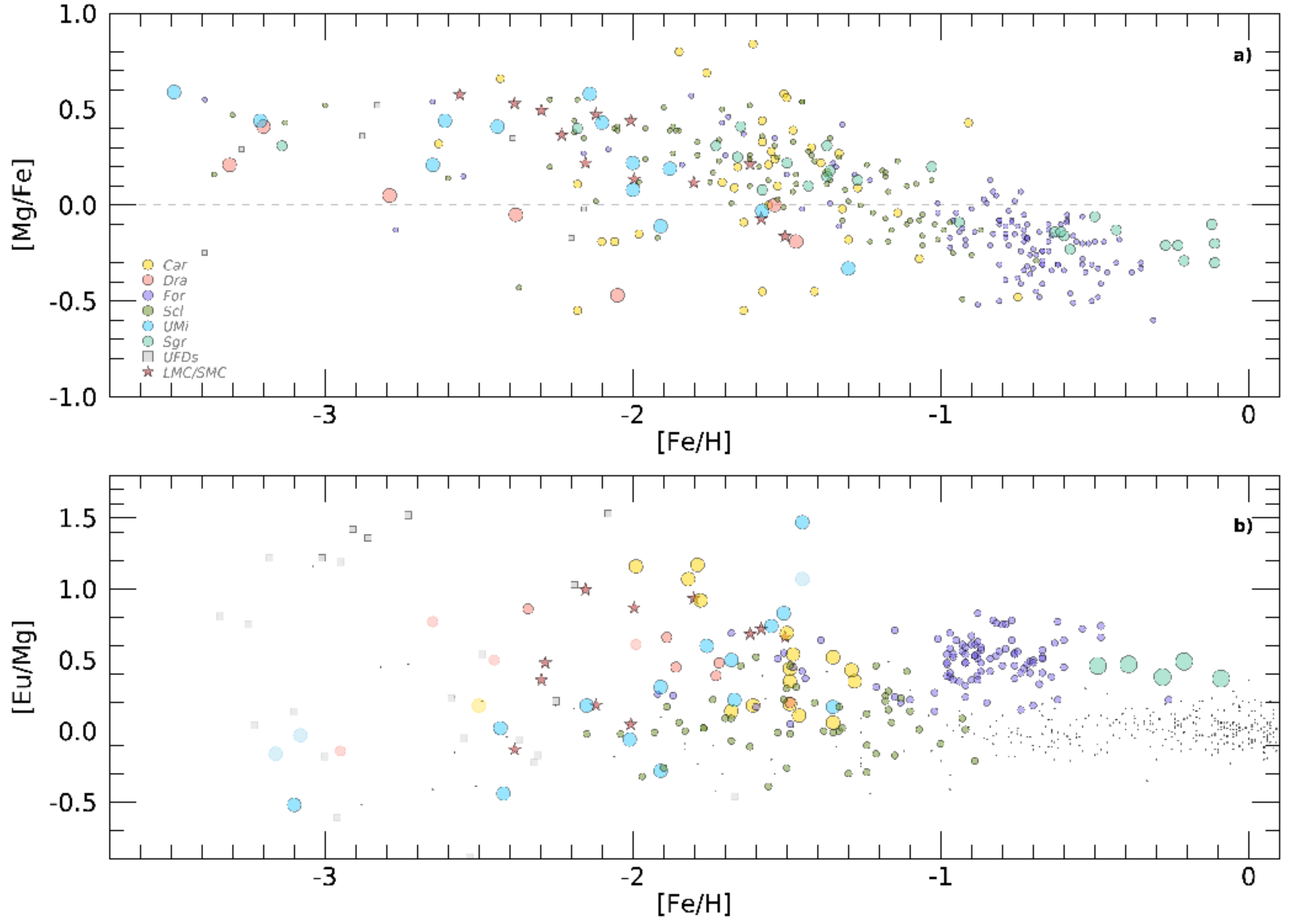}
  \caption[]{{\bf Dwarf galaxy sequences in stellar chemical abundance space.} Panel a) (top) shows [Mg/Fe] against [Fe/H] using data from \cite{Rei20}. The LMC/SMC data is from \cite{Reggiani2021}. The median metallicity uncertainty is around 0.09 dex, while [Eu/Mg] and [Mg/Fe] uncertainties are slightly smaller, but note that, naturally, there is a large variation in measurement quality between the dwarfs \citep[see][]{Rei20}. Colour represents different ETDs as shown in the legend. Dashed lines in both panels mark Solar abundance ratios. For each ETD, the [Mg/Fe] values lie on a plateau at low [Fe/H], but begin to fall later in the evolution at the knee. This characteristic behaviour is due to enrichment first by CCSNe and then later by iron-rich Type Ia SNe. Although the LMC and the SMC are the largest satellites of the MW, their knee is at lower metallicity compared to most other dwarfs, indicating unusually weak star formation in the past. Panel b) (bottom) shows [Eu/Mg] against [Fe/H] for stars in ETDs using data from~\cite{Skul2019}. Symbols without black outlines are upper limits. Small black dots are MW stars from \cite{Mishenina2013} and \cite{Venn04}. Colour-coding is the same as in the top panel. Europium is a pure r-process element. Notice the large scatter in [Eu/Mg] at metallicities below [Fe/H] $=-2$. There is evidence for trends at higher metallicity, but they vary from one ETD to the next. This may be a sign of more than one independent r-process pathway.}
   \label{fig:abundances}
\end{figure*}

Neutron-capture abundances of stars at low and intermediate metallicities in ETDs offer a powerful means of pinning down the astrophysical sites of the {\it r}-process enrichment. Of the neutron-capture elements between Ga and U, the best constraints are currently available for Sr, Y, Zr, Ba and Eu \citep[][]{Koba20}. While {\it s}-process contribution is notable in the synthesis of the first four elements, in particular at [Fe/H]$>$-2, {\it r}-process dominates the production of Eu, even at high metallicities \citep[][]{Skul2019}. Probing the relative contribution of the {\it s-} and {\it r}-process is complicated by the fact that the weak Eu absorption lines have only been accessible to high-resolution spectroscopy of bright stars in nearby dSphs \citep[][]{Bonifacio2000,Shetrone03,  Geisler2005,Cohen2009,Cohen2010,Kirby2012,McWilliam2013,Jablonka2015,Norris2017,Hansen2018,Hill2019,Skul2019,Reichert2021}. 

Until recently, it appeared that the heavy element abundances of stars in dSphs differ significantly from those in the UFDs, making the scarcity of elements such as Sr, Ba and Eu the defining characteristic of the faintest galaxies \citep[][]{Ji2019,Simon19}. This picture was overhauled with the discovery of {\it r}-process enhanced stars in an unassuming UFD Reticulum 2 \citep[][]{Ji2016,Roederer2016}. Rather than being characterised exclusively by low(er) $n$-capture enrichment at low [Fe/H], the UFDs display an extreme scatter in {\it r}-process abundance.
Subsequently, several other ETDs have been shown to contain stars enriched in heavy elements. For example, moderate {\it r}-process enhancement is reported in another UFD, Tuc 3 \citep[][]{Hansen2017}. \cite{Reichert2021} discovered several stars in the Fornax dSph with extreme Eu enrichment. They associate the birth of these objects with a recent burst of star-formation activity some 4 Gyr ago \citep[][]{deBoerFornax,Weisz14,Rusakov21}. While not at the same extreme level, significant {\it r}-process enhancement has now been registered in the metal-poor stars in the three most massive satellites, i.e. the LMC, the SMC \citep[][]{Reggiani2021} and the Gaia Sausage \citep[][]{Aguado2021,Matsuno2021}.

One way of constraining the nature of the {\it r}-process events is through timescales. The abundance ratio of Eu to a reference element produced mostly in ccSNe, for example Mg, compares the {\it r}-process delay times to those of exploding massive stars. \cite{Skul2019} examine the [Eu/Mg] ratios for the classical dSphs, UFDs as well as the Milky Way halo and disc. The scatter blows up in the metal-poor regime below [Fe/H]=-2, sampled mostly by the UFDs and the Milky Way halo. The most natural explanation for the dramatic spread in the distribution of heavy elements in UFDs is the stochasticity of their chemical evolution, where the bulk of the ISM pollution is contributed by a small number of events, perhaps even none or just one \citep[][]{Koch2008,Frebel2012}. The large variation in the heavy element enrichment and the presence of {\it r}-process enhanced stars in UFDs have been scrutinised with theoretical models. Factors such as i) dilution of the enriched gas as it propagates through the dwarf's ISM after the explosion, ii) off-center location  of the explosion (due to e.g. NS kick at birth), iii) the promptness of the star-formation episode after the explosion, may all play a role~\citep[][]{Beniamini2016,Safar2017,Tarumi2020,Jeon2021}. 

While the number of stars with well constrained {\it r}-process abundances per UFD is minuscule, the larger dSphs offer a more comprehensive picture. Some distinct patterns are noticeable in the plane of [Eu/Mg] vs [Fe/H] at metallicities above [Fe/H]=-2, as shown in the bottom panel of Fig.~\ref{fig:abundances}. For example, in Carina, [Eu/Mg] is decreasing with increasing [Fe/H], while in Ursa Minor it is growing. In Sculptor and Fornax, [Eu/Mg] shows no trend with metallicity, similar to the Milky Way, but at different levels of the abundance ratio.  [Eu/Mg] in the Magellanic Clouds stays reasonably high \citep[see][]{Reggiani2021}. In view of such variety, two independent {\it r}-process sites have been suggested, one with a short and one with a long delay time, corresponding to the ccSNs and NS mergers respectively. This is consistent with the SF and enrichment histories in the Sculptor, Fornax and Sagittarius dSphs~\citep{Skul2020} as well as the Milky Way~\citep{Mat14}. 
 
The flatness of the [Eu/Mg] ratio with metallicity in Sculptor may imply synchronisation between the {\it r}-process events and the ccSN explosions, and thus relatively short delay times for the former. This may disfavour NSs as the dominant site if their delay times are of the order of Gyrs. However, other interpretations of Sculptor's flat [Eu/Mg] chemical evolution have suggested that a single population of delayed sources alone may be sufficient~\citep{Du18,Re22}. Note that the elevated [Eu/Mg] in the Clouds around [Fe/H]$\approx-2$ is unusual, as other dwarfs such as Draco, Ursa Minor and Sculptor all show lower values. \cite{Reggiani2021} argue that the peculiar combination of [Mg/Fe] and [Eu/Mg] abundances is a sign of an extended low-level star formation from poorly enriched gas in a remote corner of the local Universe.

\section{Conclusions and Outlook}

There is a continuity of structural and kinematical properties of dwarf galaxies with increasing stellar mass, on moving from ultrafiant dwarfs (UFDs) through dwarf spheroidals (dSphs) to the gas-rich late-type dwarfs (LTDs). Star formation (SF) in the UFDs shut down early ($z\sim 3)$, continues with low-level recent activity in the brighter dSphs and remains vigorous in the LTDs today. The lowest metallicity stars in the ETDs are invaluable tracers of SF in very remote epochs, while detailed chemical abundance patterns offer clues to the nucleosynthetic history of these ancient objects. This review has stressed the variety of SF and chemical evolution histories sampled by the Milky Way's ETDs. It is precisely this diversity - once mapped out by a slew of accurate chemical abundances - that will allow us to disentangle contributions of multiple enrichment sources, simultaneously constraining the dwarfs' SFHs and the nucleosynthetic yields, thus elucidating the physics of binary evolution and supernovae explosions. This inference \cite[see][for examples]{Hasselquiest2021, Sanders21} requires large, accurate and homogeneous samples of chemical abundance measurements thus highlighting the role of wide-area high-multiplex surveys such as the ongoing GALAH~\citep{DeSilva2015} and APOGEE~\citep{Majewski2017} as well as the upcoming WEAVE~\citep{Da20}, DESI~\citep{DESI} and 4MOST~\citep{4MOST}. The heavy element production sites are also the points of feedback energy injection into the galaxy's ISM, capable of reshaping the dwarf's global morphology. However, only in the Milky Way can the effects of the internal processes be decoupled from the actions of the environment, as the satellites' orbits are pinned down by {\it Gaia}'s astrometry.

The last decade has seen substantial progress, but a number of open questions remain. First, the central parts of many ETDs are believed to be cored~\citep[][]{Wa11,Am12} with bursty star formation or outflows triggered by ccSNe responsible for transforming DM cusps into a cores~\citep{Read05,Mashchenko06,Pontzen12,Am14}. Though demonstrated as feasible in simulations, in particualr for the higher mass ETDs \citep[see][]{Bullock17}, observational confirmation of this ancient process -- a cornerstone of our theories of low-mass galaxy formation -- is lacking. However, evidence may be sought in the detection of ejected stellar material around ETDs. Wide area halo mapping with surveys like the LSST could be used to search for these faint substructures of early stars. The study of low surface-brightness outskirts is increasingly possible thanks to {\it Gaia} and narrow-band imaging. One recent example is \cite{Ch21} which finds stars out to $\sim 7$ half-light radii in Tucana III. Note however that dwarf galaxies are capable of assembling extended stellar halos around themselves via mergers with other dwarfs \citep[see e.g.][]{Deason2022}.

Secondly, the most metal impoverished stars are built from the products ejected by the first massive stars (Population III) that polluted the early Milky Way. Direct observation at high redshift of individual Population III stars is infeasible. So, the EMP stars -- together with their even poorer siblings the ultra-metal poor (UMP) stars with [Fe/H] $< -4$ -- are one of the few viable probes. Numbers of EMP and especially UMP stars are still small, despite the heroic efforts of surveys like PRISTINE~\citep{St17}. However, massive follow-up of candidates with upcoming spectroscopic surveys like WEAVE, DESI and 4MOST will soon boost the numbers substantially. Thirdly, the chemical properties of the UFDs remain enigmatic. One avenue, yet to be fully exploited, is high resolution spectroscopy of the substructures in the Milky Way halo that are the ghosts of ETDs long gone~\citep[e.g.,][]{My18}. Crucially, these ghosts are much closer than the nearest intact UFDs, which are $\gtrsim 30$ kpc away. Therefore, such disrupted substructures offer a unique window into the chemistry of very faint UFDs. This approach has been followed by \cite{Ro10} and \cite{Ag21} for the S2 stream, \cite{Ca14} for the Orphan Stream and \cite{Ha21} for the Indus Stream. There is an urgent need of more high-resolution abundance analyses of stars from intact UFDs as well, particularly for those at the metal-poor end of their metallicity distributions, in order to understand the chemical evolution and early star-forming environments of the very faintest ETDs.

This review has focused on the ETDs of the Milky Way. These may be atypical in some respects, given the quiet accretion history of the Milky Way. Studies of the SF and chemistry of the ETDs around M31 are already ongoing~\citep{Weisz19}, whilst deep {\it Hubble Space Telescope} imaging has uncovered UFDs beyond the Local Group~\citep{MP21}. The {\it Legacy Survey of Space and Time} (LSST) is predicted to increase the number of nearby ETDs by hundreds~\citep{An19}. It will also be sensitive to ETDs brighter than $M_V = -6$ in galaxy groups within $3$ Mpc~\citep{To08}. The LSST will certainly tell us whether there is a large population of ETDs still fainter than the UFDs (i.e., below the line 30 mag arcsec$^{-2}$ in Fig.~\ref{fig:sizeluminosity}), or if there exists a surface brightness threshold below which galaxies cannot form. Complementary to local studies, next generation facilities like the {\it James Webb Space Telescope} (JWST) may allow us to detect high redshift analogues of the brightest of the Local Group EDTs, assuming that their stars formed early~\citep{Avi}. The SF and enrichment histories of many more ETDs will become accessible over the next few years with the advent of extremely large telescopes with multi-object $R=5\,000-30\,000$ spectroscopy~\citep{Ji19}. High resolution spectroscopy is limited to targets within the Milky Way's virial radius and possibly M31, but medium resolution will extend to beyond the Local Group. Current facilities limit precise chemical abundances to resolved stellar populations within the Local Group. We can look forward to moving out to Mpc distances over the next decades, giving us new windows to study the detailed enrichment histories of the ETDs in different environments to the Local Group.

\section*{Declarations}
VB and NWE mapped out the scope of the review and edited the manuscript. VB provided the first draft of Sections 1 and 3, collected the data and made the Figures. NWE made the first draft of Sections 2 and 4. Both authors revised the paper in the light of the helpful referee reports. The authors declare no competing interests.

\bmhead{Acknowledgments}

The authors are indebted to Andrey Kravtsov, 
\'Asa Sk\'ulad\'ottir and Henrique Reggiani for kindly providing the data used in the Figures, as well as Matt Walker and Mike Irwin for helpful comments.

\newpage

\bibliography{references}


\end{document}